\documentclass[multphys,vecphys]{svmult}

\usepackage{makeidx}     
\usepackage{graphicx}    
\usepackage{multicol}    

\makeindex             

\newcommand\nat{Nature}
\newcommand\apj{Ap. J.}
\newcommand\apjl{Ap. J. Lett.}
\newcommand\araa{Ann. Rev. Astron. \& Astrophys.}
\newcommand\aap{Astron. \& Astrophys.}
\newcommand\mnras{MNRAS}
\newcommand\prd{Phys. Rev. D.}
\begin{document}

\title*{Gamma-Ray Bursts as Probes for
Quantum Gravity}
\author{Tsvi Piran}
\institute{Racah Institute for Physics, The Hebrew University,
Jerusalem, Israel \texttt{tsvi@phys.huji.ac.il}}
%
%
\maketitle

\section{Introduction}
\label{sec:1}

Gamma ray bursts (GRBs) are short and intense pulses of
$\gamma$-rays arriving from random directions in the sky. Several
years ago Amelino-Camelia et al. \cite{AmelinoCameliaetal98} (see
also \cite{pullin}) pointed out that a comparison of time of
arrival of photons at different energies from a GRB could be used
to measure (or obtain a limit on) possible deviations from a
constant speed of light at high photons energies. I review here
our current understanding of GRBs and reconsider the possibility
of performing these observations (see also Norris, Bonnell,
Marani, \&  Scargle \cite{Norrisetal99} for a review of the same
topic). I begin (in \S \ref{sec:1a}) with a brief discussion of
the motivation to consider an energy dependent variable speed of
light. I turn (in \S \ref{sec:2}) to a general discussion of the
detectability of deviations from a constant speed of light via
time-lag measurments. I derive constraints on the Energy range,
the distance to the sources and the needed temporal resolution of
the sources and the detectors. I then turn (in \S \ref{sec:3}) to
a short description of our current understanding of GRBs. This
section is included as a background material as for the rest of
the discussion  GRBs are just cosmological sources of high energy
photons  and we don't really care how are these photons they
produced. In \S \ref{sec:4} I return to the subject of the talk
and I describe the temporal structure and spectral properties of
GRBs. These are the key issues that are relevant for the
observations of a variable speed of light. I conclude (in \S
\ref{sec:5}) by confronting the observations needed for
determination of (or obtaining a limit on) a variable speed of
light with the properties of GRBs. I discuss some recent attempts
to obtain limits on Quantum Gravity effects
\cite{Schaefer99,Ellisetal00,Ellisetal03, Boggsetal03} and
prospects for future improvements.

\section{An Energy Dependent Speed of Light}. \label{sec:1a}

An energy dependent speed of light arises in a variety of Quantum
Gravity models, ranging from critical or noncritical string
theories, via noncommutative geometry, to canonical quantum
gravity. These models, which involve a breakdown or a modification
of Lorentz invariance at high energies, have been discussed
extensively in other lectures in this school and are reviewed
elsewhere in this volume. I focus here on a simple linear
velocity-energy relation  (see Eq. \ref{vE} below) that arises in
models for the breakup of Lorentz symmetry proposed by
Amelino-Camelia et al, \cite{AmelinoCameliaetal98}. It appears
that a similar analysis is also applicable to the case of ``DSR
deformation" of Lorentz symmetry, since the same time-of-flight
studies are considered in that
framework\cite{Amelino1,Kowalski,Magueijo,Amelino12}. In fact I
would expect that this simple linear velocity-energy relation (Eq.
\ref{vE}) would be valid, to a leading order, in many other
models.

On the phenomenological side an energy dependent speed of light
was suggested as a possible resolution of the GZK paradox
\cite{Greisen,ZatsepinKuzmin}: The observations of UHECRs   (Ultra
High Energy Cosmic Rays) above the expected (GZK) threshold for
interaction of such cosmic rays with the Cosmic microwave
background
\cite{Gonzalez,colgla,Aloisio,bertli,sato,Amelino-CameliaPiran01}.
Such energy dependence could be related to a threshold violation
at very high energies. Another possible indication for this
phenomenon is the observation of TeV photons from distant sources
\cite{Nikishov62,Gould67,Stecker92}. Such photons are expected to
be annihilated due to the interaction with the IR background.
Again threshold anomalies (that would be associated with an energy
dependent speed of light) could resolve this problem
\cite{kifu,kluz,Protheroe_Meyer,Amelino-CameliaPiran01}. In fact
Amelino-Camelia and Piran \cite{Amelino-CameliaPiran01} have
pointed out  that a simple Lorentz invariance deformation with
parameters of the order expected in various quantum gravity
theories (namely $\eta \sim 1$ in the notations used below) could
resolve both paradoxes.

\section{On the Detection of Energy Dependent Time Lags
Due to an Energy Dependent Speed of Light}. \label{sec:2}

In this short review I will not discuss the theoretical or the
phenomenological motivations for an energy dependent speed of
light. Instead I focus on the detectability of this phenomenon. I
stress that the deviations that I discuss here are drastically
different from those that arise from appearance of a photon mass.
The effects of a photon mass are most pronounced at low energies.
However, the deviations considered here depend on $E/M_{pl}$  and
are relevant only at very high energies.

Amelino-Camelia et al. \cite{AmelinoCameliaetal98} (see also
\cite{AmelinoCamelia} and other talks in this volume) pointed out
that  even a small variations in the speed of photons with
different energies could lead to observable energy dependent time
of arrival lags for photons arriving from a cosmological source.
Following Amelino-Camelia et al. \cite{AmelinoCameliaetal98}, I
consider a linear energy dependence of the form:
\begin{equation}
v = c (1 - {E \over \eta M_{pl}}) \ , \label{vE}
\end{equation}
where $M_{pl}$ is the Planck mass and $\eta$ is a dimensionless
constant. Quantum gravity effects that cause the deviation in the
speed of light are expected to take place around the Planck
energy, $ M_{pl}$. I characterize the exact energy in which these
take effect as $E_{QG}\equiv \eta M_{pl}$. The sign of $\eta$
determines the direction of these changes.

One can easily generalize the discussion and consider a more
general velocity-energy dependance,    such as: $v = c [1 - {(E /
\eta M_{pl})^\alpha}]$ \cite{Amelino-CameliaPiran01,Aloisio}.
However, for $\alpha>2$ and for the relevant energy range  and for
$\eta \approx 1$ the resulting time delays will be so short that I
don't discuss this case here.

\begin{figure}
\centering
\includegraphics[height=8cm]{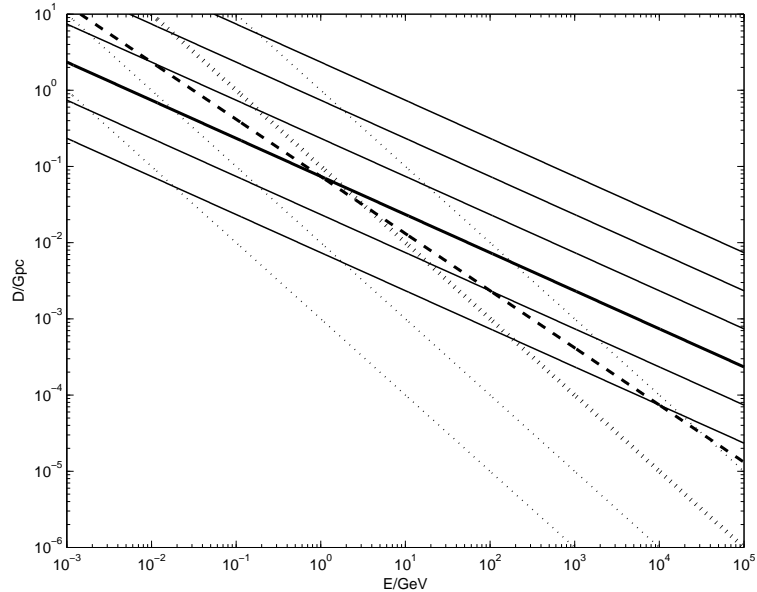}
%
%
\caption{Lines of a constant values of $\delta t_{ms} L_{50} A_4$
(solid lines) for $\delta t_{ms} L_{50} A_4 =
0.01,0.1,1,10,100,1000$. $\delta t_{ms}$ is in units of msec. The
canonical value $\delta t_{ms} L_{50} A_4 =1$ is marked by a
thicker line. Detection is possible only below a given solid line.
The single dashed line corresponds to $L(E)\propto E^{-1/2}$ and
is normalized so that $\delta t_{ms} L_{50}(1GeV)  A_4 = 1$. The
dotted lines mark lines of constant values of $\delta_{ms} \eta =
0.01,0.1,1,10$, where again $\delta t_{ms}$ is in units of msec.
The canonical value of $\delta_{ms} \eta =1$ is marked by a
thicker line. Detection is possible only above a given dotted
line. The combination of both constraints yields an allowed wedge
with a maximal distance and minimal energy. Note that the vertical
scale of distances ranges from cosmological distances at the top
($d_{Gpc}>1$) to local (galactic) distances at the bottom
($d_{Gpc}<10^{-5}$)}
\label{fig:1}       
\end{figure}

This velocity law (Eq. \ref{vE}) leads to a time lag between a
photon at energy $E$ and a very low energy photon of:
\begin{equation}
\delta t (E) \approx 10 ~{\rm msec} ~ \eta^{-1} {d_{Gpc}} E_{GeV}
\ , \label{deltat}
\end{equation}
where $d_{Gpc}$ is the distance to the source in units of Gpc and
$E_{GeV}$ is the photon's energy in GeV. Ellis et al.,
\cite{Ellisetal03} provide an exact expression as a function of
the redshift of the source. However, the approximate expression
given above is sufficient for the purpose of this work. The dotted
lines in Fig. \ref{fig:1} depict the relation between $d$ and $E$
for different values of $ \eta\delta t$. A detection, for a given
value of $\eta \delta t $,  is possible only above the
corresponding line. The value of $\delta t$  is the minimal time
delay that can be detected in the particular source.

It is clear from Eq. \ref{deltat} that we need a very high energy
source. However for these sources, because of the enormous energy
that each of the photons carries the rate of arrival of high
energy photons, $R(E)$, is very often too small. I call these
sources which are limited by  a too small rate of arrival of
photons: {\it photon starved sources}.  This has to be taken into
account as the low photon rate limits the shortest possible
detectable temporal variation as:
\begin{equation}
{1 \over R(E)} = { 4 \pi d^2 E \over A L(E) }  = 180~{\rm msec }~
{d_{Gpc}^2 E_{GeV} \over A_4 L_{50}(E) } \le  \delta t_{min} \ ,
\label{rate}
\end{equation}
where $L(E)$ is the luminosity at energy $E$ and where I have
ignored for simplicity cosmological correction factors.
$L_{50}(E)$ is the luminosity at the relevant energy interval in
units of $10^{50}$ergs/sec and $A_4$ is the area in units of
m$^2$. Again $\delta t_{min}$ is the minimal time scale that can
be detected in the particular source.

The exact limit that the combination of Eqs. \ref{rate} and
\ref{deltat} imply depends on the spectral shape,  on the overall
luminosity and on the variability time scale at the source. Quite
generally these conditions lead to an {\it upper limit} on the
distance from which the effect could be measured and to a {\it
lower limit} on the energy. As I show in \S \ref{sec:5} this limit
is important for GRBs. As an example the solid lines in Fig.
\ref{fig:1} correspond to equal values of $A L \delta t_{min}$,
for the case when the luminosity per decade is constant (i.e. the
spectral index is $-2$) and for the case that the inequality
\ref{rate} is satisfied as an equality. The dashed line on this
figure depicts the same graph for $L(E) \propto E^{-1/2}$ which is
normalized so that the luminosity per decade of energy at 1GeV is
$10^{50}$ergs/sec. A detection is possible only below these lines.
For a given combination of $\eta \delta t$ and $L A \delta t$ a
detection is possible only within a wedge outlined by the
corresponding solid line and dotted line. Namely, for a given set
of parameters there is a {\bf maximal} distance and a {\bf
minimal} energy for which the time-lag can be detected. This
suggests that in some cases (but not in the general case) a local
(galactic) source with a strong very high energy signal might be
advantageous over a weak source at a cosmological source. Indeed
this was used by Kaaret \cite{Kaaret99} to obtain a meaningful
limit on $\eta > 1.3 10^{-4}$ using the emission from the Crab
pulsar which is only at 2.2kpc

It is clear from Eq. \ref{deltat} that a cosmological distance and
a high energy are needed for a significant $\delta t$. However,
the interaction of high energy photons with the cosmic IR
background limits the distance that high energy photons can
travel. For $E\sim 100$GeV  the optical depth to $z=0.5$ is unity
\cite{Primacketal99}\footnote{Different authors make different
assumptions on the IR background and find different estimates for
the optical depth. These quantitative differences are not
important for the purpose of this work.} . Thus, we must consider
photons with $E<100$GeV. This, in turn gives an upper limit of
$\sim 6/\eta$ sec to possible magnitude of the time delay between
photons of different energies. This is independent of the source
of the emitted photons. It immediately follows that to observe
this phenomenon we need cosmological sources of $\sim$GeV photons
with a rapid and detectable variability on the time scale of
seconds or less. Amelino-Camelia et al.
\cite{AmelinoCameliaetal98}  point out that Gamma-ray bursts are
the natural candidates for this task, and indeed several groups
obtained lower limits on $\eta$ using GRBs
\cite{Schaefer99,Ellisetal00,Ellisetal03,Boggsetal03}.

\section{Gamma-Ray Bursts}
\label{sec:3}

GRBs are short and intense pulses of $\gamma$-rays that are
located at cosmological distances.  As such GRBs are ideal sources
for the effect that we are looking for. For the purpose of this
work GRBs are just a cosmological source of high energy photons.
Their  exact nature is unimportant for our ability to use the
photons to test the predictions of quantum gravity. However, it is
worthwhile, for completeness, to review briefly our current
understanding of this phenomenon. I refer the readers to several
extensive reviews
\cite{Fishman1995,P99,ParadijsARAA00,P00,Meszaros01,Hurleyetal02,%
Meszaros02a,Galama_sari,Piran2004} for more details.

It is generally accepted that GRBs are described by the
internal-external shocks model
\cite{Meszaros92,NarayanPaczynskiPiran92,ReesMeszaros94,SariPiran97}.
According to this model GRBs are produced when the kinetic energy
of an ultra-relativistic flow is dissipated. Internal shocks
within the relativistic flow produce the GRB. These shocks take
place at a distance of $\sim 10^{13}-10^{15}$cm from the center.
The short observed time scales (which violates the simple naive
rule of $\delta t < R/c$) arises because of the relativistic
motion of the flow (with a Lorentz factor $\gamma \ge 100$)
towards us. Subsequent interaction of the relativistic outgoing
flow with the surrounding matter leads to the production of an
afterglow (in x-ray, optical and radio) that lasts days, weeks,
months and in some cases even years. This takes place at distances
of $\sim 10^{16}-10^{18}$cm from the center. The flow is slowed
down due to this interaction and eventually it becomes Newtonian.

It is worthwhile to mention what is the validity of this model.
Indirect determination of the size of the afterglow of GRB 970508
\cite{Frail97} and direct measurement of the size of the afterglow
of GRB 030329 \cite{Taylor04} confirmed the predicted relativistic
motion . Additionally there is a good agreement between the
observed spectra and light curves of the afterglows and the
predictions of the relativistic shock synchrotron model. There is
also good observational evidence for the ``internal-external"
shocks transition. On the other hand, little is known about the
details of the ``inner engine" and the details how does the
collapsing core produce the required relativistic jet.

The discovery of long lasting x-ray, optical and radio afterglow
enabled the determination of the redshifts and the positions of
some bursts. The identification of bursts  within star forming
regions and the identification of Supernovae (SNe) signatures (SNe
bumps) in the afterglow of some bursts (most notably GRB 980425
and GRB 030329) revealed that long\footnote{As afterglow was seen
so far only from long burst it is not clear if short bursts are
also associated with Supernovae. In fact there are some
theoretical considerations that suggest that they are not
related.} bursts are associated with type Ic Supernovae. As the
rate of SNe Ic is much larger than the rate of GRBs it is clear
that not all Supernovae are associated with GRBs. Jet-breaks
detected in the afterglow of many bursts revealed that the bursts
are beamed into cones of a few degrees and that their total energy
is rather constant $\sim 10^{51}$ergs \cite{Frailetal01,PK01}.

The GRB-SNe association is  explained according to the Collapsar
model \cite{MacFadyen_W99}, which is a model for the ``inner
engine". According to this model a black hole - accretion disk
system forms during the core collapse. This system produces a
relativistic jet that manages to punch a hole in the supernova
envelope. The burst and the afterglow are produces along the
internal-external shocks model, once the relativistic jet has
emerged from the envelope.

\section{GRB Observations and Testing of a Variable Speed of Light}
\label{sec:4}

The possibility of observing the energy dependent time-lags depend
on four factors the distances to the sources, their temporal
structure, their spectrum and their intensity. I discuss these
three features here:

\begin{itemize}

\item{{\bf Distances:}} It is established that the bursts arise
from cosmological distances. The identified redshift record is 4.5
(GRB 000131) but it is likely that more distant bursts has been
observed but their redshift is unknown \cite{BrommLoeb}.

\item{{\bf Temporal Structure}} The bursts durations vary lasting
from a few milliseconds to a thousand seconds. The paucity of
bursts with a duration around two seconds suggest a classification
of the bursts to two groups according to their durations - long
bursts with durations longer than 2 seconds and short one with a
durations shorter than 2 seconds.

What is most important for our purpose is that most bursts show a
highly variable light curve (see for example Fig. \ref{fig:2}).
Nakar and Piran \cite{NakarPiran02}, for example analyzed the TTE
(high resolution data of the short bursts and of the first two
seconds of long ones) find in many burst sub-pulses on a time
scale of 10ms (which was about the minimal possible temporal
resolution). with sub-pulses on a scale as short as a fraction of
a millisecond \cite{Scargle98}.

\begin{figure}
\centering
\includegraphics[height=8cm]{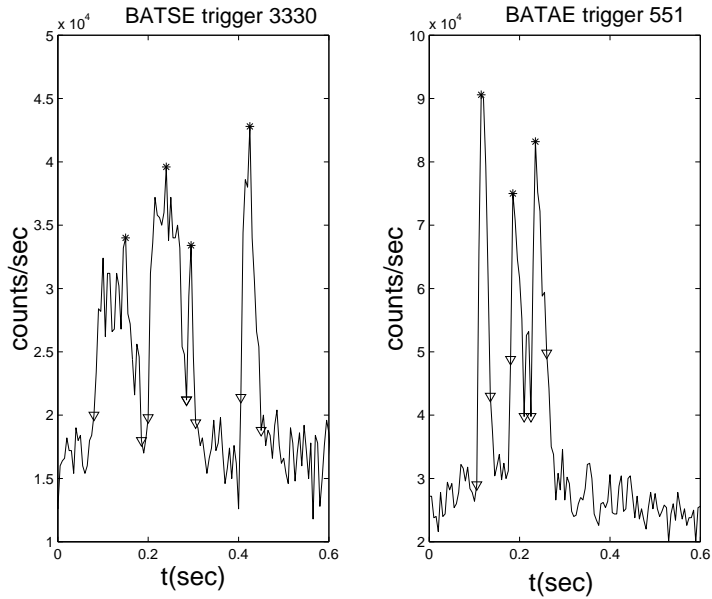}
%
%
\caption{\textbf{Left})The beginning of BATSE trigger 3330: a long
bright burst with $ T_{90}=62$sec. \textbf{Right}) The whole light
curve of BATSE trigger 551: a bright short burst with $
T_{90}=0.25$sec. The peaks are marked by stars and the triangles
mark the pulses' width. The figure demonstrates similar short time
scale structure in these bursts  (at a 5 msec resolution). From
\cite{NakarPiran02}.}
\label{fig:2}       
\end{figure}
\item{{\bf Spectrum} } The bursts's spectrum usually peaks around
a few hundred keV. Recently a subgroup of bursts, x-ray flashes,
that emits most of their energy in X-ray was discovered. In many
cases a high energy tail, with photon energies from 100 MeV to
18GeV has been observed \cite{EGRET_GRB}. The TeV detector,
Milagrito, discovered (at a statistical significance of 1.5e-3 or
so, namely at 3$\sigma$) a TeV signal coincident with GRB 970417
\cite{Milagrito_970417,Atkins03}. However no further TeV signals
were discovered so far from other 53 bursts observed by Milagrito
\cite{Milagrito_970417} or from several bursts observed by the
more sensitive Milagro \cite{Milagro_GRB}. One should recall
however, that due to the attenuation of the IR background TeV
photons could not be detected from $z>0.1$. Thus even if most GRBs
emit TeV photons those photons won't be detected on Earth.
Similarly these photons are too energetic for our purpose.

\item{{\bf Intensity}} The last factor that is important in our
consideration is the intensity of the signals. This is important
because a significant number of photons is needed to determine
exactly the timing of a pulse. The strongest observed bursts have
a fluence of $10^{-4}$ergs/cm$^2$ corresponding to $~1000$ (100keV
photons)/cm$^2$. The peak photon flux (on the
BATSE\footnote{BATSE, the Burst and Transient Source Experiment on
board on NASA's Compton-GRO, is the largest GRB detector flown so
far.} 64msec channel) is $\sim 180$photons/cm$^2$/sec.  With
typical detectors' area of  several square meters this leads to a
(100 keV range) photon rate of more than a photon per $\mu$sec
that in principle could be used to determine the temporal
structure down to a very short time scales.
\end{itemize}

The situation looks at first promising. GRBs are highly variable
bright cosmological sources providing $\gamma$-ray photons at the
right distances. Eq. \ref{deltat} reveals that energies higher
than 100MeV are needed to produce a time delay of a few
millisecond and many GRBs have such photons. At the same time many
GRBs show variability on such a time scale. However, as we see in
the next section one should proceed with cation before concluding
that GRB signals could provide a real measure of a variable speed
of light.

\section{ Caveat, Past Observations and Future Prospects}
\label{sec:5}

A careful look at the properties of GRBs uncovers, however,
problems. The main problem is that it is not clear that the high
and low energy photons seen from GRBs are emitted simultaneously.
In fact the current understanding is just the opposite. The
highest energy (18 GeV) photons discovered by EGREAT (a detector
on Compton - GRO),  were observed more than an hour after the main
burst \cite{Hurley94,Sommer94}. Similarly, when Gonzalez et al.
\cite{Gonzalez03} combined the BATSE (30keV -2Mev) data with the
EGRET data they discovered in GRB 941017 a high energy tail that
extended up to 200 MeV. This  high energy component appeared
10-20sec after the beginning of the burst and displayed a roughly
constant flux  up to 200 sec, while the main lower energy burst
decayed after several dozen seconds.

One may hope that this non-simulteneity appears only in a
``global" sense and that on a short time scale high energy photons
are emitted simultaneously with the low energy ones. While there
is not enough information on the generic time lag between very
high (100MeV and higher) and low energy (100keV) GRB photons there
is a lot of ``alarming" information on lack of simultaneity within
the BATSE band (25keV to 2MeV). Already in 1992 Fishman et al.,
\cite{Fishmanetal92} (see also Link et al., \cite{Linketal93})
noticed that the duration of GRB pulses depend on their energy and
that at lower energy the pulses are wider. Band \cite{Band97}
classifies this as a hard to soft evolution. Later Norris et al.,
\cite{Norrisetal00} noticed that this evolution corresponds to a
time lag between pulses at different energies. Typically the
higher energy pulses peak before the corresponding low energy
ones. For a sample of 174 bright bursts Norris et al.,
\cite{Norrisetal00} find typical lags between channel 1 (25-50keV)
to channel 4 (300keV-2Mev) of the order of 0.1-0.2sec with a
maximal lag of 5sec. A small fraction ($\sim 5\%$) of the bursts
have negative lags of the order of less than 0.1sec. These lags
are larger by a factor of 2-3 than the lags suggested in Eq.
\ref{deltat} for a GeV photon!

For the bursts with a known redshift Norris et al
\cite{Norrisetal00} find interesting anti-correlation between the
time lags and the peak luminosities of the bursts. This
correlation, for which there is no clear theoretical explanation,
has been used by Norris et al \cite{Norrisetal00}  to estimate the
luminosity of other bursts. It is in a general agreement with
other luminosity indicators such as the variability of the bursts
\cite{Reichartetal01}. While this correlation is not of interest
for the purpose of this work the existence  of intrinsic time lag
between photons of different energies may jeopardize the whole
prospect of detection of energy dependent time of arrival lags
arising from an energy dependent travel time. It is clear that
such an observation requires a simultaneous emission of photons at
different energies.

Ellis et al. \cite{Ellisetal03} suggest to use the redshift
dependence of the velocity induced time lags to distinguish them
from the intrinsic lags that are produced at the source. By
plotting the time lags for several BATSE bursts with a known
redshifts they obtain a limit $E_{QG}>6.9 \cdot 10^{15}$GeV or in
our notations $\eta > 6.9 \cdot 10^{-4}$. As the highest energy
photons used are of $\sim 1$MeV, this corresponds, according to
Eq. \ref{deltat} above to the conclusion that the redshift
dependent time lags are less than $\sim 0.1$sec, which is
comparable with the intrinsic time lags of these bursts
\cite{Norrisetal00}. Given the time resolution this limit ($\eta
> 10^{-3}$ seem to be (see Eq. \ref{deltat}) the best that can be done using
``low" energy ($\sim$MeV) photons.

However, there is another observational factor that appears here.
Norris et al. \cite{Norrisetal96} describe the tendency for wide
pulses to be more asymmetric, to peak later at lower energy and to
be spectrally softer, while narrow bursts are harder, more
symmetric, and nearly simultaneous. This implies that the
narrowest peaks, those that are most interesting for this
experiment have a chance of being simultaneous in both low and
high energies. Schafer \cite{Schaefer99} uses, along these lines,
the observations of one of BATSE's brightest bursts, GRB 930131
with 30keV and 80MeV photons to obtain a limit of $E_{QG}> 8.3
\cdot 10^{16}$GeV (or $\eta > 8.3 \cdot 10^{-3}$). Also along this
line Boggs et al., \cite{Boggsetal03} analyze GRB 021206. They
used the Reuven Ramaty High Energy Solar Spectroscopic Imager
(RHESSI) with an energy range of (3keV to 17MeV). They noticed
that while the lower energy ($<$2MeV) light curve of the burst is
rather irregular at higher energies the light curve exhibits a
single sharp pulse of photons extending to energies above 10MeV
with a duration of 15 msec. This enables Boggs et al.,
\cite{Boggsetal03} to set a limit of $\delta t/E  = 0.0 \pm
0.34$sec GeV$^{-1}$ from which they obtain $E_{QG}
> 1.8 \cdot 10^{17}$GeV ($\eta>0.018$). Considering Eq.
\ref{deltat} this seems to be the best that can be done with 10MeV
photons. To improve we have to get to lower temporal resolution
(which might not be possible) or to higher photon energies.

But here arises a second simple but important problem. In spite of
the fact that GRBs are the most luminous objects in the universe
at GeV energies they  are {\it photon starved}: the observed flux
is simply low. The maximal GRB fluxes at energies of a few hundred
keV are of $\sim$ 100 photons$_{100 keV}$/cm$^2$/sec. With a
several square meter detector this corresponds to a flux of $10^6$
100keV photons/sec or to a photon rate of one per $\mu$sec.
However even if GRBs emit the same energy flux at the GeV range
this flux corresponds to a meager $10^{-3}$/cm$^2$/sec GeV-photons
or to 10 GeV-photons per second with a square meter class
detector. As the minimal temporal resolution is larger than the
reciprocal of the rate of observed photons it will be impossible
to obtain a temporal resolution of better than $\sim $100msec at
the GeV range. From Eq. \ref{deltat}  this corresponds to a limit
on $\eta$ of order unity if all other problems are resolved.

The comparison of Eqs. \ref{deltat} and \ref{rate} (shown in Fig
\ref{fig:1} for a constant energy per logarithmic interval) yields
that to resolve the time lags we need a nearby ($d<1$Gpc) very
luminous GRB with a significant GeV component. Truly the rather
``small" distance will reduce $\delta t$. However, only in this
way there will be enough photons to obtain a sufficient temporal
resolution. The requirement of short distances implies that we
won't be able to use the redshift effect to distinguish between
intrinsic lags and time of flight lags. However, the fact that we
consider only very luminous bursts may resolve this problem as the
luminosity-lag correlation indicates that intrinsic lags are
smaller for more luminous bursts and they may disappear for the
very bright ones. These simple considerations are indeed supported
by the present observations.  Boggs et al., \cite{Boggsetal03}
considered a single very bright burst: GRB 0211206 which was one
of the most powerful bursts ever \cite{Nakaretal03} and was most
likely at $z \approx 0.3$, and obtained  $\eta>0.018$. This should
be compared with $\eta>0.00069$ obtained by Ellis et al.
\cite{Ellisetal03} who considered a family of weaker bursts at
cosmological distances $z\ge 1$. One has to recall however, that
such a burst occurs once per decade and it is not clear when will
the next one take place. Hopefully a suitable GRB detector will be
in orbit at that time.

The best prospect to estimate the variable velocity energy
dependent effect will be with a single observatory that could
observe both the low energy $\gamma$-rays as well as the GeV
emission. Luckily there are two planed mission that can perform
this job.

The Italian {\bf Agile} (Astro-rivelatore Gamma a Immagini
LEggero) detector \cite{Agile} is scheduled to be launched in
2005. It is a GRB detector at the energy range of 30MeV-50GeV and
a low energy detector at 10-40keV. Thus it is expected to detect
GRBs at both very high and very low energy. The temporal
resolution is about 1msec. The expected detection rate is about 10
GRBs per year at energies above 100MeV. The only limitation of
Agile is its relatively small area $\sim 0.05$m$^2$ (at 100MeV),
which might lead to a "photon starvation" problem.

An ideal observatory will be NASA's {\bf GLAST} (Gamma Ray Large
Area Space Telescope). GLAST is scheduled for launch in 2007
(Norris et al., \cite{Norrisetal99}). GLAST will include the Large
Area Telescope, LAT, which will have an effective area of 8$m^2$
and will be sensitive to photons in the 20MeV-300GeV range and
GRM, a Gamma-Ray burst Monitor which will be sensitive to photons
in the 10keV to 25MeV range.   Both the LAT and the GBM provide
the arrival time of each photon with a resolution requirements of
$<10\mu$sec (with a goal of $<10\mu$sec) and will give energies
for each detected photon.  One cannot ask for more, in terms of
the experimental design needed to study the energy dependent time
lag. Thus, if the intrinsic time lags will be resolved or shown to
be unimportant in some sub class of pulses or bursts, and this is
a very big IF in my mind,  we might be able to obtain a limit of
$\eta$ around unity towards the end of this decade.

I thank D. Band, E. Nakar and G. Amelino-Camila for helpful
remarks. This research was supported by the US-Israel Binational
Science Foundation.

%
%
%

%

\begin{thebibliography}{99.}

\bibitem{AmelinoCameliaetal98}
Amelino-Camelia, G., Ellis, J., Mavromatos, N.~E., Nanopoulos,
D.~V., \& Sarkar, S.\ 1998, \nat, 393, 763

\bibitem{pullin} R. Gambini and J. Pullin, 1999,
Phys. Rev. D59:124021

\bibitem{Norrisetal99} J.P. Norris, J.T. Bonnell, G.F. Marani, \& J.D.
Scargle, 1999, ArXiv Astrophysics e-printsScargle,
astro-ph/9912136


\bibitem{Schaefer99} Schaefer, B.~E.\ 1999,
Phys. Rev. Lett., 82, 4964



\bibitem{Ellisetal00} Ellis, J., Farakos, K.,
Mavromatos, N.~E., Mitsou, V.~A., \& Nanopoulos, D.~V.\ 2000,
\apj, 535, 139

\bibitem{Ellisetal03} Ellis, J., Mavromatos, N.~E.,
Nanopoulos, D.~V., \& Sakharov, A.~S.\ 2003, \aap, 402, 409

\bibitem{Boggsetal03}
Boggs, S.~E., Wunderer, C.~B., Hurley, K., \& Coburn, W.\ 2003,
ArXiv Astrophysics e-prints, astro-ph/0310307

\bibitem{AmelinoCamelia}
Amelino-Camelia, G., 2004, {\it This Volume}.

\bibitem{Amelino1} G. Amelino-Camelia, 2002,
Int. J. Mod. Phys. D11:35-60
\bibitem{Kowalski} J. Kowalski-Glikman, S. Nowak,
2003, Int. J. Mod. Phys. D12:299-316
\bibitem{Magueijo} J.Magueijo and L. Smolin, 2003, Phys. Rev. D67:044017

\bibitem{Amelino12} G., Amelino-Camelia, L. Smolin, A.
Starodubtsev, 2004, Classical \& Quantum Gravity 21, 3095

\bibitem{Greisen} K.~Greisen. 1996,   Phys. Rev. Lett., 16:748
\bibitem{ZatsepinKuzmin} G.~T. Zatsepin \& V.~A. Kuzmin, 1966,
Sov.  Phys.-JETP Lett., 4:78

\bibitem{Gonzalez} L. Gonzalez-Mestres, 1997, physics/9704017

\bibitem{colgla} S.~Coleman, S.L.~Glashow, 1999,
Phys.~Rev.~D59. 116008.

\bibitem{Aloisio} R. Aloisio, P. Blasi, P.L. Ghia, A.F. Grillo,
2000, Phys. Rev. D62:053010

\bibitem{bertli} O.~Bertolami and C.S.~Carvalho, 2000,
Phys.~Rev.~D61, 103002.

\bibitem{sato} H.~Sato,
astro-ph/0005218.

\bibitem{Amelino-CameliaPiran01}
Amelino-Camelia, G.~\& Piran, T.\ 2001, \prd, 64, 036005

\bibitem{Nikishov62} Nikishov A.I., 1962, {Sov. Phys. JETP}, {14}, 393

\bibitem{Gould67} Gould J., Schreder G., 1967, {Phys. Reports}, {155.5}, 1404

\bibitem{Stecker92} Stecker F.W., De Jager O.C. and Salomon M.H.,
1992, \apjl, {390}, L49

\bibitem{kifu} T.~Kifune,
\apjl 1999, 518, L21.

\bibitem{kluz}
Klu{\' z}niak, W.\ 1999, Astroparticle Physics, 11, 117


\bibitem{Protheroe_Meyer} , R.~J.~\&
Meyer, H.\ 2000, Physics Letters B, 493, 1



\bibitem{Kaaret99} Kaaret, P.\ 1999, \aap, 345, L32



\bibitem{Primacketal99} Primack, J.~R., Bullock, J.~S.,
Somerville, R.~S., \& MacMinn, D.\ 1999, Astroparticle Physics,
11, 93

\bibitem{Fishman1995} Fishman, G.~J., and C.~A. Meegan,  1995, \araa 33, 415.

\bibitem{P99} Piran, T., 1999, Phys. Reports 314, 575.
\bibitem{ParadijsARAA00} van Paradijs, J., C.~Kouveliotou, and
  R.~A.~M.~J. Wijers,  2000, \araa 38, 379.

\bibitem{P00} Piran, T.,
  2000, Phys. Reports, 333, 529.
\bibitem{Meszaros01} M{\' e}sz{\' a}ros, P.,
  2001, Science,  291, 79-84

\bibitem{Hurleyetal02} Hurley, K.,R.~Sari, and
  S.~G. Djorgovski, 2002, astro-ph ???

\bibitem{Meszaros02a} M{\' e}sz{\' a}ros, P.,
  2002, \araa 40, 137.

\bibitem{Galama_sari}
Galama, T., and R. Sari,2002, in Relativistic Flows in
  Astrophsyics, LNP 589, edited by Axel Guthmann et al., Springer-Verlag,
  Berlin,  123.
\bibitem{Piran2004} Piran, T.,
  2004, Rev. Modern Phys. in press, astro-ph/0405503.

\bibitem{Meszaros92} Meszaros, P.~\& Rees,
M.~J.\ 1992, \mnras, 257, 29P
\bibitem{NarayanPaczynskiPiran92} Narayan,
R., Paczynski, B., \& Piran, T.\ 1992, \apjl, 395, L83


\bibitem{ReesMeszaros94} Rees, M.~J.~\&
Meszaros, P.\ 1994, \apjl, 430, L93

\bibitem{SariPiran97} Sari, R.~\& Piran, T.\
1997, \apj, 485, 270


\bibitem{Frail97} Frail, D.~A., Kulkarni,
S.~R., Nicastro, S.~R., Feroci, M., \& Taylor, G.~B.\ 1997, \nat,
389, 261



\bibitem{Taylor04} Taylor, G.~B., Frail, D.~A., Berger, E., \& Kulkarni, S.~R.\ 2004,
ArXiv Astrophysics e-prints, astro-ph/0405300

\bibitem{Rhoads99} Rhoads, J.~E.\ 1999, \apj, 525,
737


\bibitem{Frailetal01} Frail, D.~A., et al.\
2001, \apjl, 562, L55

\bibitem{PK01}  Panaitescu, A.~\& Kumar, P.\ 2001, \apjl, 560, L49
\bibitem{MacFadyen_W99}
MacFadyen A.~I., and S.~E. Woosley, 1999, \apj, 524,
 262,


\bibitem{BrommLoeb} Bromm, V.~\& Loeb, A. 2002, \apj, 575, 111

\bibitem{NakarPiran02} Nakar, E.~\& Piran, T.\
2002, \mnras, 330, 920



\bibitem{Scargle98} Scargle,
J.~D., Norris, J.~P., \& Bonnell, J.~T.\ 1998, American Institute
of Physics Conference Series, 428, 181

\bibitem{EGRET_GRB} Dingus, B.~L.~\&
Catelli, J.~R.\ 1998, Abstracts of the 19th Texas Symposium on
Relativistic Astrophysics and Cosmology, held in Paris, France,
Dec.~14-18, 1998.~Eds.: J.~Paul, T.~Montmerle, and E.~Aubourg (CEA
Saclay).,

\bibitem{Milagrito_970417} Atkins, R., et al.\
2004, \apjl, 604, L25

\bibitem{Atkins03} Atkins, R., et al.\
2000, \apjl, 533, L119

\bibitem{Milagro_GRB} McEnery, J.\ 2002, APS Meeting
Abstracts, A3007

\bibitem{Hurley94} Hurley, K.\ 1994, \nat, 372, 652

\bibitem{Sommer94} Sommer, M., et al.\ 1994, \apjl, 422, L63

\bibitem{Gonzalez03} Gonz{\' a}lez,
M.~M., Dingus, B.~L., Kaneko, Y., Preece, R.~D., Dermer, C.~D., \&
Briggs, M.~S.\ 2003, \nat, 424, 749



\bibitem{Fishmanetal92} Fishman, G.~J., Meegan,
C.~A., Wilson, R.~B., Horack, J.~M., Brock, M.~N., Paciesas,
W.~S., Pendleton, G.~N., \& Kouveliotou, C.\ 1992, American
Institute of Physics Conference Series, 265, 13



\bibitem{Linketal93} Link,
B., Epstein, R.~I., \& Priedhorsky, W.~C.\ 1993, \apjl, 408, L81


\bibitem{Band97} Band, D.~L.\ 1997, \apj, 486, 928

\bibitem{Norrisetal00} Norrisf,
J.~P., Marani, G.~F., \& Bonnell, J.~T.\ 2000, \apj, 534, 248


\bibitem{Reichartetal01} Reichart, D.~E., Lamb,
D.~Q., Fenimore, E.~E., Ramirez-Ruiz, E., Cline, T.~L., \& Hurley,
K.\ 2001, \apj, 552, 57

\bibitem{Nakaretal03} Nakar, E.,
Piran, T., \& Waxman, E.\ 2003, Journal of Cosmology and
Astro-Particle Physics, 10, 5

\bibitem{Agile} M. Tavani et al., 2004, Astron. Astrophys. Suppl. Ser. 138, 569-570

\bibitem{Norrisetal96} Norris, J.~P., Nemiroff,
R.~J., Bonnell, J.~T., Scargle, J.~D., Kouveliotou, C., Paciesas,
W.~S., Meegan, C.~A., \& Fishman, G.~J.\ 1996, \apj, 459, 393





\end{thebibliography}
%



\printindex
\end{document}